\documentclass[8pt]{article}  \usepackage{times}
\usepackage{graphicx}

\usepackage{color}

\usepackage[round,numbers,sort&compress]{natbib} 

\newcommand{\be}{\begin{equation}}
\newcommand{\ee}{\end{equation}}

\newcommand{\pd}[2]{\frac{\partial #1}{\partial #2}} 
 
\newcommand{\pdc}[3]{\left( \frac{\partial #1}{\partial #2}
 \right)_{#3}} 
\let\baraccent=\= 
\renewcommand{\=}[1]{\stackrel{#1}{=}} 

\newcommand{\eref}[1]{Eq.~(\ref{#1})}

\begin{document}

\twocolumn[{\LARGE \textbf{Fluctuations of systems in finite heat reservoirs with applications to phase transitions in lipid membranes\\*[0.2cm]}}
{\large Lars D. Mosgaard, Andrew D. Jackson and Thomas Heimburg$^{\ast}$\\*[0.1cm]
{\small $^1$Niels Bohr Institute, University of Copenhagen, Blegdamsvej 17, 2100 Copenhagen \O, Denmark}\\

{\normalsize ABSTRACT\hspace{0.5cm} In an adiabatically shielded system the total enthalpy is conserved. Enthalpy fluctuations of an arbitrarily chosen subsystem must be buffered by the remainder of the total system which serves as a heat reservoir. The magnitude of these fluctuations depends on the size of the reservoir. This leads to various interesting consequences for the physical behavior of the subsystem. As an example, we treat a lipid membrane with a phase transition that is embedded in an aqueous reservoir. We find that large fluctuations are attenuated when the reservoir has finite size.  This has consequences for the compressibility of the membrane since volume and area fluctuations are also attenuated.

We compare the equilibrium fluctuations of subsystems in finite reservoirs with those in periodically driven systems. In such systems, the subsystem has only finite time available to exchange heat with the surrounding medium. A larger frequency therefore reduces the volume of the accessible heat reservoir.  Consequently, the fluctuations of the subsystem display a frequency dependence. 

While this work is of particular interest for a subsystem displaying a transition such as a lipid membrane, some of the results are of a generic nature and may contribute to a better understanding of relaxation processes in general.
\\*[0.0cm] }}
]

\noindent\footnotesize {$^{\ast}$corresponding author, theimbu@nbi.dk}\\


\normalsize
\section*{Introduction}

The enthalpy fluctuations of an adiabatically shielded system are zero by definition.  The enthalpy of arbitrary subsystems contained within the total system can only fluctuate by the exchange of heat with the rest of the system which we call `the reservoir'. In a simple homogeneous system this leads to temperature fluctuations in both the subsystem and the `reservoir' that are trivially related and that depend only on the size of the two parts of the system. An example would be enthalpy and temperature fluctuations in a small water volume that is embedded into a larger water reservoir of finite size. One can also consider cases where the subsystem is of different physical nature than the reservoir. Such a subsystem could be a particular vibrational mode in a macromolecule that couples to the rest of the molecule that serves as a reservoir. One may also consider subsystems that are spatially separated from the reservoir, e.g., macromolecules or membranes dissolved in an aqueous buffer.  The purpose of this paper is to treat this problem in all generality and apply it to the particularly interesting case of a subsystem that can undergo a phase transition while embedded in a homogeneous medium that displays no transition. In particular, we discuss the case of a lipid membrane with a melting transition when the membrane is in contact with a finite aqueous volume that serves as a heat reservoir.

When varying temperature, lipid membranes display cooperative melting transitions in which both enthalpy and entropy of the individual molecules change at a melting temperature, $T_m$ \cite{Heimburg2007a}. At this temperature, the heat capacity has a maximum. According to the fluctuation-dissipation theorem, at constant temperature the heat capacity is proportional to the enthalpy fluctuations of the membrane and closely related to the fluctuation time-scales.

Heat capacity is typically measured in a differential scanning calorimeter (DSC). A DSC controls the temperature very precisely and records the heat absorbed by the sample when the temperature is changed. Therefore, the temperature of the reservoir is fixed by the instrumental setup, which is intended to behave like an infinite reservoir with constant temperature.  In finite adiabatic systems (with constant total enthalpy), however, the temperature of the reservoir is not constant because it exchanges heat with the membrane due to fluctuations. Consequently, there are fluctuations of the reservoir temperature that are completely correlated with the enthalpy fluctuations of the subsystem (here, the membrane).  Thus, the temperature of the reservoir is only constant on average with fluctuations that can be either large or small depending on the size of the reservoir.  In this publication we show that the size of the (water) reservoir has a significant effect on the magnitude of the fluctuations and the relaxation time scales of the subsystem (the lipid membrane).  

There have been very few attempts to model systems in a finite reservoir \cite{Creutz1983, Milchev1994}, and these are of limited generality and not applicable to the lipid membrane system. The lipid membrane is distinct from many other systems due to its pseudo two-dimensional nature. While the membrane is effectively two-dimensional, it is embedded in a three-dimensional reservoir with which it can exchange heat. The overall system thus consists of two coupled systems with a total enthalpy that is constant but fluctuating for each of the two sub-systems.  Here, we present a statistical mechanics framework for modeling the lipid melting transition in a finite heat reservoir, i.e., a membrane in a very small water volume.  This problem is of more than academic interest.  

The heat capacity $c_p$ is an equilibrium property of a system and therefore does not possess a timescale.  When a system is probed for finite times (or when the system is driven by an external periodic force), it may not be possible to establish equilibrium with the entire reservoir.  Such non-equilibrium systems can be approximated by an equilibrated adiabatic system consisting of the membrane and a reservoir of finite size.  Adiabaticity ensures that the total enthalpy fluctuations of this combined system are precisely zero.  The fluctuation-dissipation theorem cannot be used to calculate the heat capacity, and other methods must be used.  It is, however, possible to calculate the enthalpy fluctuations for the membrane alone.   In the limit of large reservoirs, these fluctuations describe the usual equilibrium heat capacity.  For smaller reservoirs unable to support large enthalpy fluctuations, the fluctuations in the enthalpy of the membrane will necessarily be reduced.  Such effects should be most pronounced near the maximum of the equilibrium heat capacity.  It is very important to point out that this argument holds for all fluctuations of extensive quantities such as volume and area of the subsystem, which are closely related to the enthalpy fluctuations. Therefore, our considerations can be extended to the elastic properties of the subsystem that are determined by the volume and area fluctuations. Our analysis contains a reinterpretation of the adiabatic compressibility. 

We note that some authors \cite{Nielsen1996b} have performed calculations in systems driven externally at a well-defined frequency to determine a ``dynamic heat capacity'' or ``frequency dependent heat capacity''.  The authors (Nielsen \& Dyre) suggest that the frequency dependent heat capacity can be understood as an equilibrium property of the system. In the limit of an arbitrarily small frequency, which corresponds to an infinite reservoir, this dynamic heat capacity is identical to the usual equilibrium heat capacity.   For finite frequencies, it is closely related to the enthalpy fluctuations of membranes in finite size reservoirs studied here using Monte Carlo simulations.  We discuss our finding of reservoir-size dependent membrane fluctuations in the context of the frequency dependence of the heat capacity of membranes determined in periodic perturbation experiments \cite{vanOsdol1989, vanOsdol1991a} and with the frequency dependence of sound \cite{Halstenberg2003, Mosgaard2012}. Our findings suggest a close connection between the frequency dependence of both the compressibility and the sound velocity of membranes and the size of the available water reservoir.


\section*{Theory}\label{Theory}

\subsection*{Fluctuations in finite reservoirs}\label{model}

\begin{figure}[!h]
	\centering
		\includegraphics[width= 0.8 \linewidth]{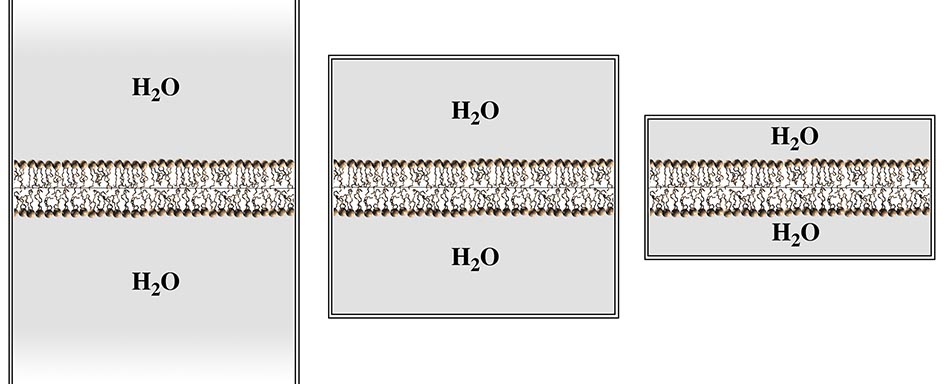}
	\parbox[c]{8cm}{ \caption{\textit{Three scenarios for a lipid membrane subsystem in an aqueous reservoir: Left: The membrane is embedded in an infinite water reservoir with constant temperature. Center and Right: The membrane is embedded in a finite size water reservoir. The total system consisting of membrane and water is adiabatically shielded. Thus, enthalpy fluctuations of the membrane now are coupled to both fluctuations in enthalpy and temperature of the water reservoir. }
	\label{Figure0}}}    
\end{figure}

Enthalpy is strictly conserved in an adiabatically insulated system.  Any heat released or absorbed by a subsystem must be exchanged with the surrounding system which we call the reservoir. Consequently, the properties of the reservoir will also fluctuate. Typically, one considers the fluctuations of a small system in an infinite heat reservoir (for the example of a membrane embedded into an aqueous reservoir see Fig. \ref{Figure0}, \textit{left}) that effectively keeps the temperature of the reservoir constant. This is also the situation in calorimetric experiments.  In such an infinite system, temperature fluctuations of the reservoir vanish.  This is not the case for a finite system (Fig. \ref{Figure0}, \textit{right}), where care is required to guarantee that the enthalpy is strictly conserved.  As shown below, this implies that the temperature of the reservoir fluctuates in correlation with fluctuations of the subsystem. \\

textbf{The Gibbs free energy change associated with a state change in the subsystem is }
\be
\Delta G_s=\Delta H_s-T\Delta S_s	\;,
\label{gw1b}
\ee 
where the index `s' denotes the subsystem. 
During this change in state, heat is transferred from the subsystem to the reservoir. 

The free energy change of the reservoir $\Delta G_r$ (the index `r' denoting the reservoir) upon the absorption of the heat $\Delta H_r=-\Delta H_s$ is given by 
\be
\Delta G_r = \Delta H_r - T \Delta S_r,
\label{gw1}
\ee
where $\Delta H_r$ is the change in enthalpy of the reservoir and $\Delta S_r$ is the associated entropy change in the reservoir. $T$ is the temperature, and $\Delta G=\Delta G_s+\Delta G_r$ the free energy change of the total system. If the reservoir absorbs heat from a fluctuation of the subsystem, the change in the enthalpy of the reservoir is naturally fixed to exactly this amount since the total system conserves enthalpy. 

From the local fluctuations of temperature, the change in the reservoir's entropy associated with the transfer of enthalpy internally between the two sub-ensembles can be calculated as follows: 
\be
c_{p}^r=T \pdc{S_r}{T}{P}  \quad \Rightarrow \quad\Delta S_r  =  \int^{T_r^b}_{T_r^a} \frac{c_{p}^r}{T}dT,
\ee
where $c_{p}^r$ is the heat capacity of the reservoir and $\Delta S_r$ is the corresponding change in entropy. 
The heat capacity of the reservoir is assumed to be constant. The reservoir temperature $T_r$ before the change in the state of the subsystem is defined as $T_r^a$ and after the change as $T_r^b$ (with $\left<T_r^a\right> = \left<T_r^b\right> =T$, averaged over  time. $T$ is the constant temperature of the total system that enters the Boltzmann factors). The entropy change of the reservoir is then given by
\be
\Delta S_r = c_{p}^r \ln \frac{T_r^b}{T_r^a},
\label{s_r}
\ee
where $(T_r^b-T_r^a)$ is the temperature change of the reservoir associated to absorbing a given amount of heat, $\Delta H_r$. Since  $c_p^r\left(T_r^b-T_r^a\right)=\Delta H_r$ for constant $c_p^r$,  the temperature $T_r^b$ of the reservoir after absorbing $\Delta H_r$ is given by:
\be
T_r^b = \frac{\Delta H_r}{c_p^r} + T_r^a\;.
\label{temp_step}
\ee

Using \eref{s_r}, \eref{gw1} can be rewritten as
\begin{eqnarray}
\Delta G_r &=& \Delta H_r - T c_p^r \ln \frac{T_r^b}{T_r^a} \nonumber \\
&=& \Delta H_r - T c_p^r \ln \left( \frac{\Delta H_r / c_p^r + T_r^a}{T_r^a}\right) \;.
\label{gw}
\end{eqnarray}

Note that in the limit $c_p^r \rightarrow \infty$ the free energy $\Delta G_r \rightarrow 0$ independently of the magnitude of $\Delta H_r$.

\subsection*{The probability of a state change in a finite reservoir}
We can now determine the acceptance probability of a change in the state of the subsystem in a finite adiabatic system. It is given by
\begin{eqnarray}
p= \frac{K}{1+K}	\qquad;\qquad K=\exp\left(-\frac{\Delta G_s+\Delta G_r}{RT}\right)\;,
\label{acc}
\end{eqnarray}
which obeys detailed balance. If it is decided to allow a change of state of the subsystem during a Monte-Carlo simulation, the enthalpy associated with this change is absorbed or supplied by the reservoir. $T_r^a$ of the reservoir is updated to the value of $T_r^b$. 

Since $\Delta H_s+\Delta H_r=0$, the equilibrium  is completely governed by entropy differences:
\begin{eqnarray}
\Delta G&=&-T(\Delta S_s+\Delta S_r)\nonumber\\
&=&-T\left(\Delta S_s + c_p^r \ln \left( 1-\frac{\Delta H_s }{c_p^r T_r^a}\right)\right)	\;.
\label{Gtot}
\end{eqnarray}

In the limit of $c_p^r\rightarrow \infty$, $\Delta G\rightarrow\Delta G_s$, as expected. In this limit, the fluctuations of the subsystem are independent of the nature of the reservoir. It is also obvious that for finite $c_p^r$ there is a maximum fluctuation that can be carried by the system: $\Delta G \rightarrow \infty$ for $\Delta H_s\rightarrow c_p^r\cdot T_r^a$. For vanishing reservoir size, no enthalpy fluctuations in the subsystem are possible.

It is important to point out that the result of these considerations are general. In any physical system, the probability of heat transfer from any arbitrarily chosen subsystem `s' to a reservoir `r' consisting of the rest of the total system is a function of the heat capacity of the reservoir.

\section*{Modeling lipid membrane fluctuations in a finite aqueous reservoir}
Below, we apply these concepts to the fluctuations in lipid membranes embedded into an aqueous reservoir. In particular, we consider the case of the cooperative melting transition from an ordered gel to a disordered fluid membrane. 

Monte Carlo simulations have frequently been used to analyze the cooperative behavior of membranes. Some early applications can be found in \cite{Doniach1978, Mouritsen1983, Sugar1994, Heimburg1996a}.   Enthalpy fluctuations are the central element in such simulations. The parameters for the simulation are the melting enthalpies and entropies of the lipid components and the nearest neighbor interactions. The overall temperature is assumed to be constant and identical to that of the aqueous reservoir.  The enthalpy fluctuates during the simulation.  The heat capacity at constant pressure can be calculated from the enthalpy fluctuations and yields $c_p=(\left<H^2\right>-\left<H\right>^2)$ $/RT^2$, where $\left<...\right>$ denotes the statistical average and $T$ is the (constant) temperature of the reservoir.  The fluctuation relation can easily be calculated from a canonical ensemble of N identical systems that are allowed to exchange heat. Due to ergodicity, the time evolution of a single system at absolutely constant temperature leads to the same distribution of states. The latter can be studied in Monte Carlo simulations, and it is meaningful to determine the heat capacity of a membrane from the fluctuations observed in such simulations.  

The assumption of constant reservoir temperature and the resultant neglect of reservoir temperature fluctuations are only permissible if the size of the reservoir is infinite. In a finite reservoir, the separation of the membrane from its surroundings is not permissible because the enthalpy fluctuations of the membrane and of the reservoir are correlated. Nevertheless, considering the fluctuations of the membrane alone can provide meaningful insights into the behavior of a membrane. 
The Gibbs free energy of each configuration of the lipid subsystem consisting of N lipids is given by:
\be
G_s=G_g+N_f (\Delta H-T\Delta S)+N_{gf}\omega_{gf} \; ,
\label{doniach1}
\ee
where $G_g$ denotes the Gibbs free energy of the ground state (with all lipids in the ordered gel state). $\Delta H$ and $\Delta S$ are the molar excess enthalpy and entropy of the melting transition, which can be obtained from the calorimetric experiment. $N_f$ is the number of lipids in the fluid state, $N_{gf}$ is the number of unlike nearest neighbor contacts associated with an interfacial enthalpy contribution. The parameter $\omega_{gf}$ describes unlike nearest-neighbor interactions and is typically positive. It is responsible for the cooperativity of the transition, i.e., the half width of the melting transition and the size of domains in the transition regime.

We further assume that each lipid is associated with $N_{water}$ water molecules with which the membrane exchanges heat during the simulation. Further, the lipid chains possess a heat capacity, $c_p^{chain}$, which is due to vibration within the molecular bonds. This heat capacity is also part of the heat reservoir. Thus, the total heat capacity per lipid of the reservoir, $c_p^r$, is given by 
\be
c_{p}^r=N_{water}\cdot c_p^{water} +c_p^{chain}\;.
\label{cp_reservoir1}
\ee
This number has to be multiplied by the total number of lipids to obtain the total heat capacity of the reservoir. For more details and parameter values see App. \ref{MC-simulations}.


\section*{Results}

\subsection*{Simulations of a lipid membrane in a finite heat reservoir}
\begin{figure}[!h]
	\centering
		\includegraphics[width= 1 \linewidth]{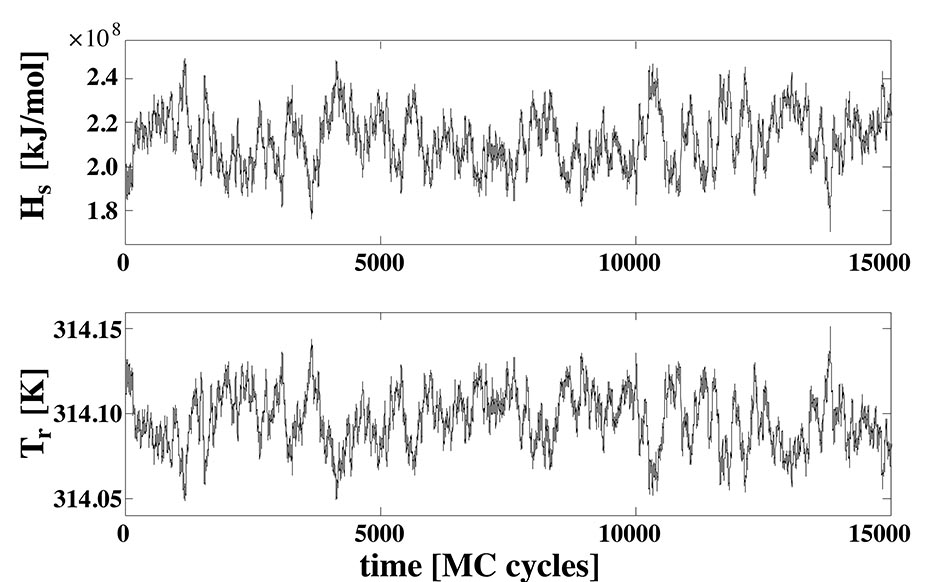}
	\parbox[c]{8cm}{ \caption{\textit{Traces of membrane enthalpy, $H_s$, and reservoir temperature, $T_r$, from Monte Carlo simulations (100 $\times$ 100 matrix). Top: Fluctuations in enthalpy $H_s$ of a lipid membrane with 1000 water molecules associated to each lipid. The enthalpy is given for the total lipid matrix (molar units). Bottom: Temperature fluctuations in the aqueous reservoir. The water molecules serve as a reservoir for the heat released from the membrane. The membrane enthalpy and the temperature are correlated due to the adiabatic boundary conditions. $\Delta H_s$ and $\Delta T_r$ are exactly proportional functions.}
	\label{Figure1}}}    
\end{figure}

We first consider the effect of the finite heat reservoir on the lipid melting transition. In order to illustrate the coupling between the membrane enthalpy $H_s$ and the reservoir temperature $T_r$, we performed a Monte Carlo simulation at the melting temperature (314.05 K) with 1000 water molecules per lipid.  This is shown in Fig. \ref{Figure1}.   Due to eq. (\ref{temp_step}), $H_s$ and $\Delta T_r$ are exactly proportional functions. 

\begin{figure}[!h]
	\centering
		\includegraphics[width= 1 \linewidth]{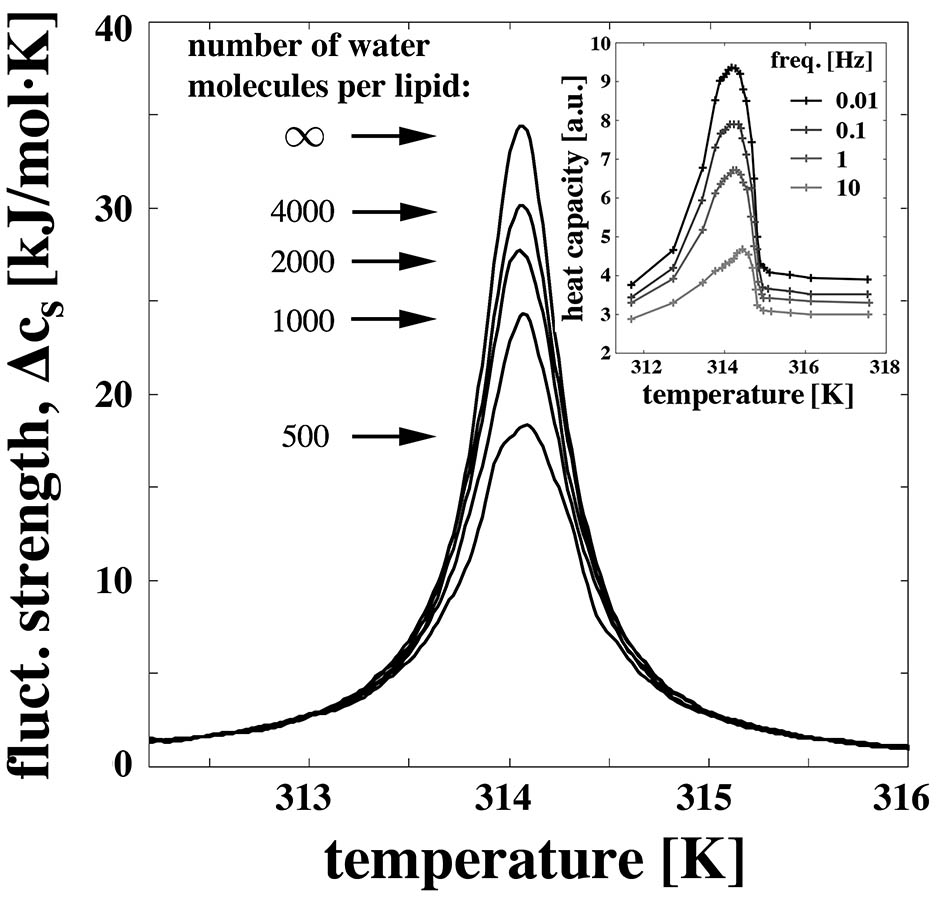}
	\parbox[c]{8cm}{ \caption{\textit{Fluctuation strength, $\Delta c_s$, of the lipid membrane for five different sizes of associated water reservoirs. The isothermal limit corresponds to an infinite number of water molecules per lipid. The curves have been smoothed by cubic spline fitting. Error bars have been omitted for clarity (cf. error bars in Figs. \ref{Figure3} and \ref{Figure4}). The inset shows frequency dependent heat capacities, $c_p(\omega)$, measured by van Osdol et al. (adapted from \cite[]{vanOsdol1991a}).}
	\label{Figure2}}}    
\end{figure}

Subsequently, we calculated the fluctuations of the enthalpy of the membrane and determined the function $\Delta c_s=\left(\left<\Delta H_s^2\right>-\left<\Delta H_s\right>^2\right)/RT^2$, which we call the fluctuation strength of the membrane.  In Fig. \ref{Figure2}, it is shown close to the transition temperature. We show the $c_s$-profiles for five different sizes of the aqueous reservoir: 500, 1000, 2000, 4000 and an infinite number of H$_2$O molecules per lipid. The later case corresponds to the isothermal limit, i.e., to the heat capacity $\Delta c_p$ of the membrane. It can be seen that  a reduction of the size of the available heat reservoir also reduces the fluctuation strength $\Delta c_s$ of the lipid membrane. This lowering is due to the suppression of large enthalpy fluctuations in the lipid membrane. 
In the limit of infinite reservoirs, the excess heat integrated over the melting transition is given by $\Delta H=\int c_p dT$. For finite reservoirs, however, $\int c_s dT < \Delta H$. For this reason, we do not call $c_s$ a heat capacity.

The dependence of the fluctuation strength on reservoir size is also shown in Fig. \ref{Figure4} for 4 different temperatures close to the transition temperature. Fig. \ref{Figure2} shows that the position and width of the fluctuation function profile in the melting transition are unaltered, meaning that the depletion of the fluctuation strength with smaller heat reservoirs occurs without broadening the transition (Fig. \ref{Figure2}). For comparison, the inset of Fig. \ref{Figure2} shows experimental data for frequency-dependent heat capacities from van Osdol and collaborators adapted from \cite[]{vanOsdol1991a}. The relation between finite size systems and frequency dependence is considered in the Discussion section.\\
In order to demonstrate the robustness of our approach we show in Appendix \ref{System-size} that these results are independent of the overall system size as long as the number of water molecules per lipid is constant.

\subsection*{Fluctuation timescales in finite systems}

\begin{figure}[!htb]
	\centering
		\includegraphics[width= 1 \linewidth]{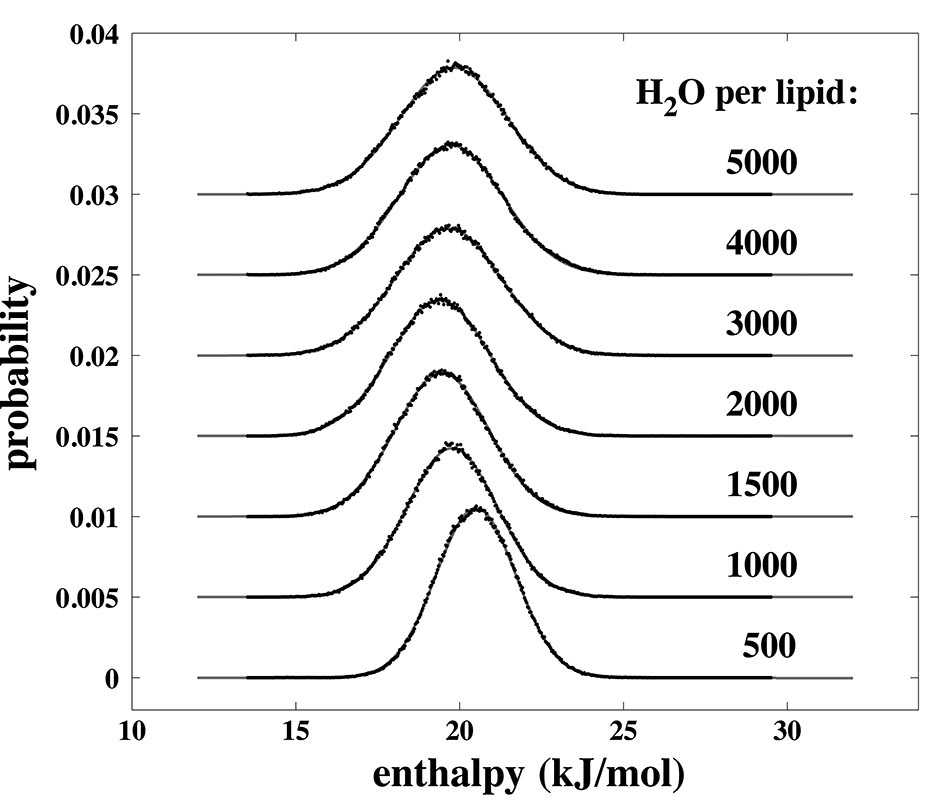}
	\parbox[c]{8cm}{ \caption{\textit{Probability distribution of enthalpy states close to the transition maximum for different reservoir sizes. The simulated distribution (symbols) is well described by a Gaussian distribution with a half width that is closely related to the fluctuation strength.
}
	\label{Figure5}}}    
\end{figure}

Fig. \ref{Figure5} shows the probability distribution of enthalpy fluctuations close to the transition temperature for various reservoir sizes. It can be seen that the distributions are Gaussian,
\be
P(H_s)=\frac{1}{\sqrt{2\pi \sigma^2}}\exp\left(-\frac{(H_s-\left<H_s\right>)^2}{2\sigma^2}\right)
\label{probability-distr}
\ee
where the variance of the fluctuations, $\sigma^2=\left<H_s^2\right>-\left<H_s\right>^2$, is directly related to the fluctuation strength ($c_s=\sigma^2/RT^2$).
Following Einstein \cite{Einstein1910}, this implies that entropy fluctuations of the system are harmonic with 
\be
S(H_s)\approx -R\frac{(H_s-\left<H_s\right>)^2}{2\sigma^2} + \mbox{const.}
\label{entropy-potential}
\ee
with an entropy maximum at $H_s=\left<H_s\right>$.
The use of linear response theory allows us to conclude that, for a fixed reservoir size, the relaxation behavior of enthalpy fluctuations is described by a single exponential with a relaxation time constant, $\tau$, given by \cite{Grabitz2002, Seeger2007}
\be
\tau=\frac{T^2}{L}\Delta c_s \;,
\label{relaxation-time}
\ee
where $L$ is a phenomenological coefficient setting the absolute time scale of the cooperative processes.
This relation implies that the relaxation times in our simulations are directly proportional to the magnitude of the fluctuations. We find this also in a direct correlation analysis of the cooperative enthalpy fluctuations in the simulation (not shown, see \cite{Grabitz2002} for examples). Smaller reservoir sizes result in a reduced fluctuation strength with a smaller fluctuation time constant, i.e., fluctuations are faster.  We will discuss this feature in the context of frequency dependent heat capacities in the Discussion section.

\subsection*{Linking the effective heat capacity to the adiabatic compressibility}
We now consider some consequences of the above results concerning the magnitude of volume or area fluctuations of the membrane in finite reservoirs, and their relation to the adiabatic compressibility. The results are especially instructive if the reservoir is a nearly incompressible medium such as water while the subsystem displays large volume or area fluctuations such as those shown by membranes close to transitions.

The specific isothermal area compressibility (i.e., infinite reservoir) is given by
\be
\kappa_{T}^{A}=-\frac{1}{A}\left(\pd{A}{\Pi}\right)_T \;,
\label{kappa_T_1}
\ee
where $\Pi$ is the lateral pressure and $A$ is the membrane area. Close to the melting transition, the isothermal compressibility can be approximated by
\be
\kappa_{T}^{A}\approx \kappa_{T,0}^A+\frac{\gamma_{A}^2 T}{A}\Delta c_{p} \;,
\label{kappa_T_2}
\ee
where $\Delta c_p$ is the excess heat capacity \cite{Heimburg1998, Pedersen2010}. In \eref{kappa_T_2} we used the experimentally found relation $\Delta A = \gamma_A \Delta H$, with $\gamma_A=0.89$ m$^2$/J for a lipid bilayer of DPPC \cite{Heimburg1998, Heimburg2005c}. 
The adiabatic area compressibility is related to the isothermal compressibility and is given by \cite{Wilson1957}
\be
\kappa_{S}^{A} \equiv -\frac{1}{A}\left(\pd{A}{\Pi}\right)_S=\kappa_{T}^{A}-\frac{T}{A \; c_{p}^{system}}\left(\pd{A}{T}\right)_{\Pi}^2 \;.
\label{kappa_a}
\ee
This relation has been derived for equilibrium systems using the Maxwell relations \cite{Wilson1957}.  Here, $c_{p}^{system}$ is assumed to be the heat capacity of the total thermodynamic system, i.e., the excess heat capacity of the  lipid membrane, $\Delta c_p$, plus the heat capacity of the reservoir, $c_p^{r}$ (lipid chains and aqueous buffer),
\be
c_{p}^{system}=\Delta c_p+ c_p^{r} \;.
\label{cp-system}
\ee 
Assuming that $(\partial A/\partial T)_{\Pi}$ in the lipid melting transition region is completely dominated by the change in area associated with the transition, we obtain \cite{Halstenberg1998}:
\begin{eqnarray}
\kappa_{S}^{A} &\approx& \kappa_{T,0}^A+\frac{\gamma_{A}^2 T}{A}\Delta c_{p}-\frac{\gamma_{A}^2 T}{A}\frac{\Delta c_{p}^2}{c_{p}^{\mbox{\tiny system}}} \nonumber \\
& = & \kappa_{T,0}^A+\frac{\gamma_{A}^2 T}{A}\Delta c_p\cdot \left(1-\frac{\Delta c_{p}}{c_{p}^{\mbox{\tiny system}}}\right).
\label{kappa_a_a}
\end{eqnarray}
It is easily seen that the term in brackets approaches unity when the heat capacity of the total system is much larger than the excess heat capacity of the lipid membrane.  This implies that the adiabatic and isothermal compressibilities of the membrane are equal for a very large reservoir. 
\begin{figure}[!b]
	\centering
		\includegraphics[width= 1 \linewidth]{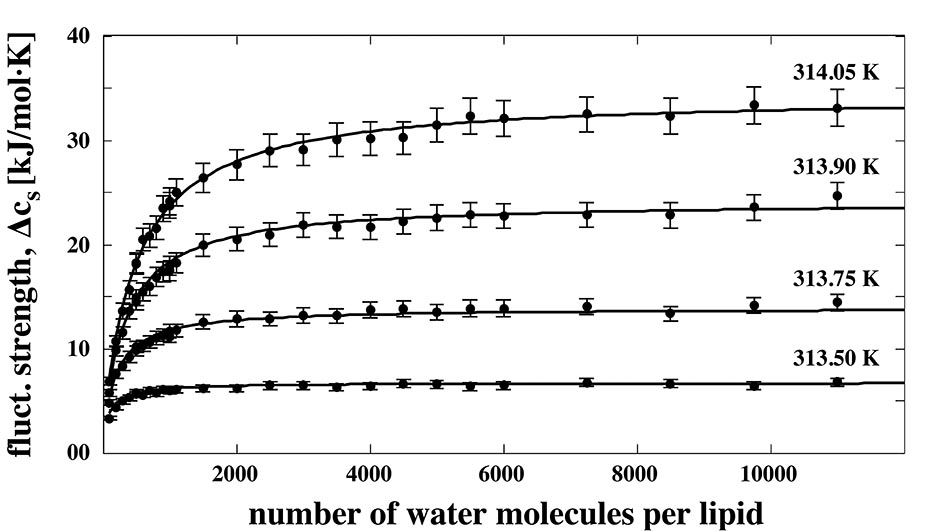}
	\parbox[c]{8cm}{ \caption{\textit{Verification of the analytical Ansatz. The effective heat capacity calculated as a function of reservoir size calculated from eq. \eref{cp-finite} (solid lines) and the fluctuation strength, $\Delta c_s$, from the simulations (symbols) at four different temperatures. The analytical formalism yields a very good approximation of the simulated data. }
	\label{Figure4}}}    
\end{figure}

Following Halstenberg \emph{et al.} \cite{Halstenberg1998}, we postulate that the effective heat capacity of the lipid membrane in a finite size reservoir is given by
\be
\Delta c_p^{\mbox{\tiny eff}}=\Delta c_{p}\cdot \left(1-\frac{\Delta c_{p}}{c_{p}^{\mbox{\tiny system}}} \right)\;.
\label{cp-finite}
\ee
with an associated adiabatic compressibility of
\be
\kappa_S^A=\kappa_{T,0}^A+\frac{\gamma_A^2 T}{A}\Delta c_p^{\mbox{\tiny eff}}\;,
\label{kappa-finite}
\ee
which is formally similar to eq. (\ref{kappa_T_2}). The treatment for the isothermal and adiabatic volume compressibilities is absolutely analogous.

In order to test whether this is a reasonable definition of the membrane heat capacity, it is therefore interesting to compare the above heat capacity with the fluctuation strength of the membrane, $\Delta c_s$, obtained from the Monte Carlo simulations (Fig. \ref{Figure2}). In the Monte Carlo simulation, the heat capacity of the total heat reservoir, $c_p^r$, is an input parameter. The excess heat capacity of the lipid melting transition in the isothermal case is known, because it corresponds to the standard Monte-Carlo simulation with constant reservoir temperature \cite{Sugar1994}. We can therefore calculate the effective heat capacity analytically from eq. (\ref{cp-finite}) and compare it with the simulation results.  Fig. \ref{Figure4} shows the fluctuation strength of the membrane from Monte-Carlo simulation as a function of reservoir size (symbols) at four different temperatures. Due to the linear relation between fluctuation strength and fluctuation time scales discussed above, the time scales display the same dependence on reservoir size. The solid lines show the analytical calculation from eq. (\ref{cp-finite}). Within the estimated error, perfect agreement between \eref{cp-finite} and the simulated fluctuation strength was found, indicating that these are identical functions: $\Delta c_p^{\mbox{\tiny eff}}=\Delta c_s$. 

Our results also indicate that the isothermal and the adiabatic compressibility are not fundamentally different functions. They merely reflect different sizes of the available heat reservoir. They are equally related as the heat capacity and the fluctuation strength in finite reservoirs as seen from eqs. (\ref{kappa_a_a}) and (\ref{cp-finite}).


\section*{Discussion}
Here, have shown that the enthalpy fluctuations of an arbitrary part (subsystem) of an adiabatically insulated total system (total enthalpy is constant) depends on the entropy of the total system, i.e., it depends on the combined entropy of the subsystem and the reservoir. This entropy can be regarded as a harmonic potential which depends on the relative size of subsystem and reservoir (i.e., the rest of the total system). Linear response theory then leads to interesting connections between enthalpy fluctuations of the subsystem, its fluctuation lifetimes and its adiabatic compressibility. While many of our considerations are general, we have applied them to the special case of lipid membranes surrounded by an aqueous reservoir. The fact that enthalpy, volume and area fluctuations of lipid membranes are proportional functions \cite{Heimburg1998} allows us to find very simple relations between seemingly different thermodynamic response functions.

In calorimetric experiments, membranes (in the form of a dispersion of vesicles) are coupled to an aqueous reservoir and the calorimeter itself. It is generally assumed that the calorimeter serves as an infinite heat bath guaranteeing a constant temperature of the reservoir. If the temperature of the reservoir is absolutely constant, it is meaningful to assign a heat capacity $c_p$ to a subsystem, and the integral $\int_{T_1}^{T_2} c_p dT=\Delta H$ yields the enthalpy change of the subsystem upon a variation of the temperature. We have shown that this is not the case for a finite reservoir that necessarily has temperature fluctuations that are intimately coupled to the enthalpy fluctuations of the subsystem.

The mean square fluctuations of two systems cannot be added when they are correlated, and it is not meaningful to assign heat capacities to individual parts of the total system. However, one can consider enthalpy fluctuations of subsystems that we called the `fluctuation strength' $\Delta c_s$ of the membrane.  For finite size reservoirs, it is generally true that $\Delta c_s < \Delta c_p$. The integral of $c_s$ over temperature does not yield the enthalpy difference of the system at different temperatures. For this reason, we do not call $c_s$ a heat capacity.

\subsection*{Frequency dependent heat capacity and the relation to the finite reservoir}
Experimentally, it is hard to test the dependence of the membrane fluctuations on the aqueous volume directly because at very low water content the phase diagrams of lipid membrane dispersions change. However, one can consider frequency-de\-pendent processes where only a short time is available for the membrane system to exchange heat with the buffer. Under such circumstances, only a small volume of the aqueous buffer can contribute as a reservoir. As a result, the size of the volume that communicates with the membrane is frequency-dependent.

In periodic perturbation experiments one can determine the amplitude of the periodic heat uptake. This function has often been called the `frequency-dependent' or `dynamic heat capacity', $c_p(\omega)$. This term has been coined in analogy with the definition of the equilibrium heat capacity $dQ/dT$. However, in periodic perturbation experiments both $dQ$ and $dT$ display a dependence on frequency. $c_p(\omega)$ is a complex function with an amplitude and a phase shift between $dQ(\omega)$ and $dT(\omega)$. This phase shift is absent at zero frequency. There are basically two ways of determining the frequency dependent heat capacity. The first consists of a periodic temperature variation imposed on the system from the outside, which is linked to a periodic uptake and release of heat, such as described by \cite{Mayorga1988}. The second method consists of a periodic variation of pressure of an adiabatically shielded volume. The observable is the periodic variation in reservoir temperature \cite{Johnson1983}. The frequency dependent heat capacity is determined indirectly using the Clausius-Clapeyron equation. What is actually observed  in the case of lipid membranes is the transfer of heat from the membrane to the reservoir \cite{vanOsdol1989, vanOsdol1991a}. This situation is in fact comparable to our case that considers temperature fluctuations in the reservoir generated by enthalpy fluctuations in the membrane. For this reason we compared the frequency dependent heat capacity by \cite{vanOsdol1991a} with the fluctuation strength in finite reservoirs (Fig. \ref{Figure2}). The inset of Fig. \ref{Figure2} shows the results of these experiments on DPPC vesicles for four frequencies between 0.01 to 10 Hz. They display a striking similarity to our simulations when varying reservoir size in two respects: 1. The half width of the excess heat capacity profile is unchanged but its amplitude decreases when increasing frequency or decreasing reservoir size. 2. The effect on amplitude is most pronounced in the transition, because the fluctuation time scales are much larger due to critical slowing-down.

In contrast to the enthalpy fluctuations, the equilibrium heat capacity does not possess an intrinsic time scale. Nielsen and Dyre \cite{Nielsen1996b} have thoroughly  analyzed the frequency dependent heat capacity and its coupling to fluctuation relations. They define $c_p(\omega)$ as the fraction of the equilibrium fluctuations with time scales shorter than $\tau=1/2\pi\omega$.  Clearly, $c_p(\omega)$ captures only those equilibrium fluctuations that are faster than the characteristic time scale of the oscillation. In other words, it captures those heat transfer processes that have equilibrated within the time $t < \tau$.  In the limit of $\omega\rightarrow 0$, the frequency dependent heat capacity $c_p(\omega)$ therefore approaches the equilibrium heat capacity, $c_p$. Our present simulation considers heat transfer into a finite reservoir in an equilibrium situation. The reduction in reservoir size attenuates the large fluctuations. By demonstrating the Gaussian nature of the fluctuations, we have also shown that fluctuation relaxation is single exponential with a time scale related to the size of the reservoir (Fig. \ref{Figure5}). Thus, relaxation of heat into a finite reservoir resembles the relaxation of heat in finite time as discussed above. 

Consider a membrane embedded in an infinite water reservoir (Fig.\ref{Figure0}, left) that is subject to periodic variation of the lateral pressure applied to the membrane.  It is reasonable to assume the this will lead to an exchange of heat with an adjacent layer of water that is finite due to the finite time scale for heat transport in water. In the first phase of the perturbation, heat is released into the aqueous layer; in the second phase it is reabsorbed. The volume of the contributing water layer is likely to be directly related to the timescale of the oscillation. 
 
In the past, we have demonstrated for lipid membranes that the equilibrium volume and area fluctuations are directly proportional to the enthalpy fluctuations \cite{Heimburg1998, Ebel2001} as are the relaxation times following temperature and pressure perturbations. This suggests a proportionality between equilibrium heat capacity and isothermal volume or area compressibility. The adiabatic compressibility is also an equilibrium property that can be derived from isothermal properties by using Maxwell relations. It is not intuitive why the concept of an adiabatic compressibility can successfully be used for describing dynamic or frequency dependent phenomena. While the frequency dependent heat capacity is not a thermodynamic function, we have shown here that one can nevertheless draw a reasonable analogy between a properly defined `frequency dependent heat capacity' and a `frequency dependent compressibility' and suggest a proportional relationship for the two.  In analogy to eq. (\ref{cp-finite}), one can also postulate that the frequency dependent excess heat capacity of the membrane assumes the following form:
\be
\Delta c_p^{\mbox{\tiny eff}}(\omega)=\Delta c_{p}\cdot \left(1-\frac{\Delta c_{p}}{c_{p}^{\mbox{\tiny system} }(\omega)} \right)\;,
\label{cp-omega}
\ee
where $c_p^{system}(\omega)$ is the reservoir size accessible in the finite time $\tau=1/2\pi \omega$. The excess adiabatic compressibility is then given by
\be
\Delta \kappa_S(\omega) =\frac{\gamma_A^2 T}{A}\Delta c_p^{\mbox{\tiny eff}}(\omega)\;.
\label{kappa-omega}
\ee
If $c_{p}^{\mbox{\tiny system}}\rightarrow \infty$, the frequency dependent heat capacity approaches the equilibrium excess heat capacity, and the adiabatic compressibility approaches the isothermal compressibility.  Understanding the timescale of heat transfer from the membrane subsystem into the aqueous reservoir might help formulating dispersion relations.

While eq. (\ref{kappa-omega}) expresses a tentative rather than a derived form of the frequency dependence of the compressibility, it has been used successfully in describing the ultrasonic frequency dependence of the 3-dimensional sound velocity of lipids in the Mhz regime. Halstenberg et al. \cite{Halstenberg1998} performed experiments on DPPC vesicles using a resonator with a frequency of 7.2 MHz, which corresponds to a timescale that is much faster than that of the cooperative domain size fluctuation in equilibrium. The speed of sound in the volume is given by 
\be
c=\sqrt{\frac{1}{\kappa_S^V\rho^V}}
\label{sound}
\ee
where $\kappa_S^V$ is the adiabatic volume compressibility and $\rho^V$ is the mass density. The experimentally measured speed of sound of lipid dispersions was correctly predicted by assuming that the heat capacity of the lipid chains is dominant at such high frequencies.  Again, the rationale is that there is insufficient time for heat to diffuse into the aqueous volume at these frequencies.  

The frequency dependence of sound is called `dispersion' and is of considerable importance for sound propagation phenomena in matter. We have previously proposed that electromechanical solitons with strong similarities to the action potential can propagate in biomembranes and nerves \cite{Heimburg2005c, Heimburg2007b, Andersen2009, Lautrup2011, Villagran2011}.  Such solitons are a consequence of the simultaneous presence of non-linear elastic constants and dispersion close to melting transitions.  Although many details remain to be understood, we have also shown that the dispersion relation is related to the thermodynamic behavior of membranes \cite{Mosgaard2012}.  In particular, the dispersion relation sets a natural timescale for the propagating nerve pulse. Similarly, the fluctuation time scales correspond to the typical open-time of lipid channels \cite{Wunderlich2009, Gallaher2010}. It seems likely that the time scale of fluctuations is of significant biological relevance. 


\section*{Conclusion}
We have constructed a framework for modeling the fluctuations of arbitrary subsystems embedded in an adiabatically shielded reservoir. This method was applied to the lipid melting transition in a finite adiabatically insulated aqueous reservoir. We show that the magnitude of the cooperative fluctuations of the membranes depends on the size of the associated  reservoir. As a consequence, the elastic constants of the membrane also depend on reservoir size. It seems plausible to compare this effect to frequency dependent measurements where only parts of the environment of a membrane can contribute as a reservoir for the heat transfer. We believe that the present considerations may contribute to the better understanding of relaxation processes in general and the dispersion relation of lipid membranes that is important for setting the time scale of dynamic processes such as nerve pulse propagation. \\

\textbf{Acknowledgments:}
This work was supported by the Villum foundation (VKR 022130).

\begin{appendix}\label{Methods}
\section{Monte Carlo simulations}\label{MC-simulations}

We have modeled the melting transition of a single lipid membrane using the Doniach model \cite{Doniach1978}, which is a modified version of the Ising model with two lipid states, gel and fluid, instead of two spins. This differs from the Ising model in that the two lipid states are not only different in enthalpy but also in entropy.  This is due to the higher degeneracy of states of each lipid molecule in the fluid phase.  We used Monte Carlo simulations employing the Glauber algorithm for the individual simulation steps \cite{Glauber1963}. Such simulations are described in detail by \cite{Sugar1994, Heimburg1996a}.

Simulations were typically carried out on a triangular  lattice with 100 $\times$ 100 sites with periodic boundary conditions. Each lattice point represents one lipid which can either be in the gel or the fluid state. All simulations were equilibrated for at least 30 times the correlation time before sampling, effectively meaning more than $6\cdot10^4$ Monte Carlo cycles at the transition maximum. The equilibration was carried out by assuming a constant water bath temperature in the first step. In a second step we considered finite reservoir size using an algorithm described below.  In analogy to the heat capacity, we defined the excess fluctuation strength $\Delta c_s=\left(\left<\Delta H_s^2\right>-\left<\Delta H_s\right>^2\right)/RT^2$  that we calculated from the excess enthalpy fluctuations of the lipid membrane (enthalpy $H_s$) embedded into the finite reservoir. The statistical error was estimated using the Jackknife Method.  We emphasize that the fluctuation strength  $\Delta c_s$ is identical to the equilibrium excess heat capacity defined as $\Delta c_p=(dQ/dT)_p$ only in the limit of infinite reservoirs and constant reservoir temperature.

In the present simulation we used the following parameters for modeling the heat capacity profiles of dipalmitoyl phosphatidylcholine (DPPC) large unilamellar vesicles (LUV):  $\Delta H =36400 \;J/mol$ (melting enthalpy), $\Delta S=115.9 \; J/mol \cdot K$ (melting entropy) and $\omega_{gf}=1326.0 \; J/mol$ \cite{Ivanova2001} leading to a melting temperature of $T_m=314.05 \;K$ and a transition half width of about 1 K. The heat capacity of water was taken to be $c_p^{water}=75 \; J/K  \cdot mol$  which corresponds to the value of 1 cal/g$\cdot$ K for free water. The heat capacity of the chains was set to $c_p^{chain}=1600 \; J/K \cdot mol$ which was determined experimentally by Blume \cite{Blume1983} for gel state DPPC. The total heat reservoir is shared by all lipids in the lipid membrane. The minimum number of water molecules per lipid considered in any simulation is $100$. 

The simulated heat capacity profiles and the estimated statistical errors were smoothed using cubic spline fits. 

\section{System size dependence}\label{System-size}
\begin{figure}[!htb]
	\centering
		\includegraphics[width= 1 \linewidth]{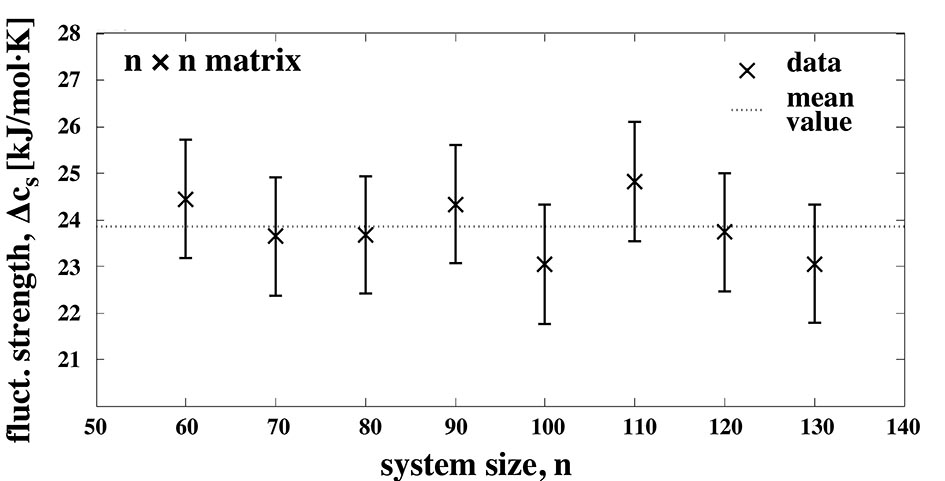}
	\parbox[c]{8cm}{ \caption{\textit{Fluctuation strength, $\Delta c_s$, at the transition temperature $T_m$ of the lipid membrane in a finite system with $1000$ water molecules per lipid.  The simulation was performed for different sizes, $n$,  of the lipid membrane. A system size of $n$ denotes an  $n\times n$ matrix. The fluctuation strength per lipid is independent of system size within the error of the calculation.}
	\label{Figure3}}}    
\end{figure}

Fig. \ref{Figure3} shows that the calculated fluctuation strength (per mole of lipid) is independent of the total number of lipids assuming a fixed reservoir size  per lipid (here 1000 H$_2$O molecules per lipid)  within statistical error. This behavior was demanded in the Theory section and demonstrates the robustness of our approach. 

\end{appendix}

\small{


}
\end{document}